\documentclass[12pt,twocolumn]{article}

\usepackage{arxiv}
\usepackage{amsmath}
\usepackage{graphics}
\usepackage[utf8]{inputenc} % allow utf-8 input
\usepackage[T1]{fontenc}    % use 8-bit T1 fonts
\usepackage{url}            % simple URL typesetting
\usepackage{booktabs}       % professional-quality tables
\usepackage{amsfonts}       % blackboard math symbols
\usepackage{nicefrac}       % compact symbols for 1/2, etc.
\usepackage{microtype}      % microtypography
\usepackage{lipsum}

\usepackage{amsfonts}
\usepackage{amssymb}
\usepackage{bbold}
\usepackage{booktabs}
\usepackage{ulem}
\usepackage{xr}
\usepackage[justification=centering ]{subcaption}
\usepackage{graphicx}
\usepackage{amsmath}
\usepackage[usenames,dvipsnames]{xcolor}
\usepackage{multirow}
\usepackage{lscape}
\usepackage{hyperref}

\usepackage[sorting = none, backend = biber, style=phys]{biblatex}
\usepackage{color}
\usepackage{tikz}
\usetikzlibrary{shapes}

\usepackage{bm}
\usepackage{amsmath,amssymb}
\renewcommand{\vec}[1]{\boldsymbol{#1}}
\definecolor{minered}{HTML}{FF0000}
\definecolor{mineblue}{HTML}{4169E1}
\definecolor{minedarkblue}{HTML}{AFEEEE}
\definecolor{minegreen}{HTML}{228B22}
\definecolor{colorCa01}{HTML}{FA9070}
\definecolor{colorCa015}{HTML}{F37457}
\definecolor{colorCa02}{HTML}{EC583E}
\definecolor{colorCa03}{HTML}{CF2613}
\definecolor{colorCa04}{HTML}{7A180C}
\definecolor{colorCa045}{HTML}{511008}
\definecolor{colorCa055}{HTML}{000000}

\newcommand{\Ca}{\text{Ca}}
\newcommand{\Rey}{\text{Re}}

\newcommand{\cs}{c_{\mbox{\scriptsize s}}}

\newcommand\corr[1]{\textcolor{black}{#1}}

\title{\Large Droplet dynamics in homogeneous isotropic turbulence with the immersed boundary-lattice Boltzmann method}

\author{Diego Taglienti$^{1}$\And Fabio Guglietta$^{1,\dagger}$\And Mauro Sbragaglia$^{1}$\\ %
\and
$^{1}$Department of Physics \& INFN, Tor Vergata University of Rome, Via della Ricerca Scientifica 1, 00133, Rome, Italy\\
\and
$^{\dagger}$\texttt{fabio.guglietta@roma2.infn.it}
}

\begin{document}
\twocolumn[
  \begin{@twocolumnfalse}
    \maketitle
    \begin{abstract}
      We develop a numerical method for simulating the dynamics of a droplet immersed in a generic time-dependent velocity gradient field. This approach is grounded on the hybrid coupling between the lattice Boltzmann (LB) method, employed for the flow simulation, and the immersed boundary (IB) method, utilized to couple the droplet with the surrounding fluid. We show how to enrich the numerical scheme with a mesh regularization technique, allowing droplets to sustain large deformations. The resulting methodology is adapted to simulate the dynamics of droplets in homogeneous and isotropic turbulence, with the characteristic size of the droplet being smaller than the characteristic Kolmogorov scale of the outer turbulent flow. We report on statistical results for droplet  deformation and orientation, collected from an ensemble of turbulent trajectories, as well as comparisons with theoretical models in the limit of small deformation.
    \end{abstract}
  \end{@twocolumnfalse}
]

%%%%%%%%%%%%%%%%%%%%%%%%%%%%%%%%%%%%%%%
\section{\label{sec:intro}Introduction}
%%%%%%%%%%%%%%%%%%%%%%%%%%%%%%%%%%%%%%%%
Understanding the dynamics of droplets in turbulence is fundamentally important~\cite{Mashayek2003} and practically relevant for a broad spectrum of applications including mixing and blending processes~\cite{Sanfeld2008,Tang1994,Yukuyama2015}, petroleum industry~\cite{Windhab2005,Kumar2021}, environmental remediation~\cite{Li1998,Hendraningrat2014}, food industry~\cite{TabiloMunizaga2005,Lee2013}, and hemodynamic engineering~\cite{blood2003}. The characterization of droplet features such as elongation and orientation in such scenarios is challenging: the flow inside and outside the droplet couple in a non-linear way at the droplet interface, and this coupling is further enriched by the complexity brought by time-dependent strain rates exerted by the flow surrounding the droplet. Analytical approaches are only possible in the limiting case of small deformations and/or with limiting assumptions on the properties of the flow~\cite{taylor1932viscosity,taylor1934formation,rallison1980note,SARKAR2001}. The use of numerical simulations is therefore of paramount importance to understand the dynamics of such droplets~\cite{milan2018lattice,Fuster2009,elghobashi2019direct,Roccon2017,Mukherjee2019,CrialesiEsposito2022,CrialesiEsposito2023,Deberne2024}.

\begin{figure*}[th!]
\centering
\includegraphics[width=.9\linewidth]{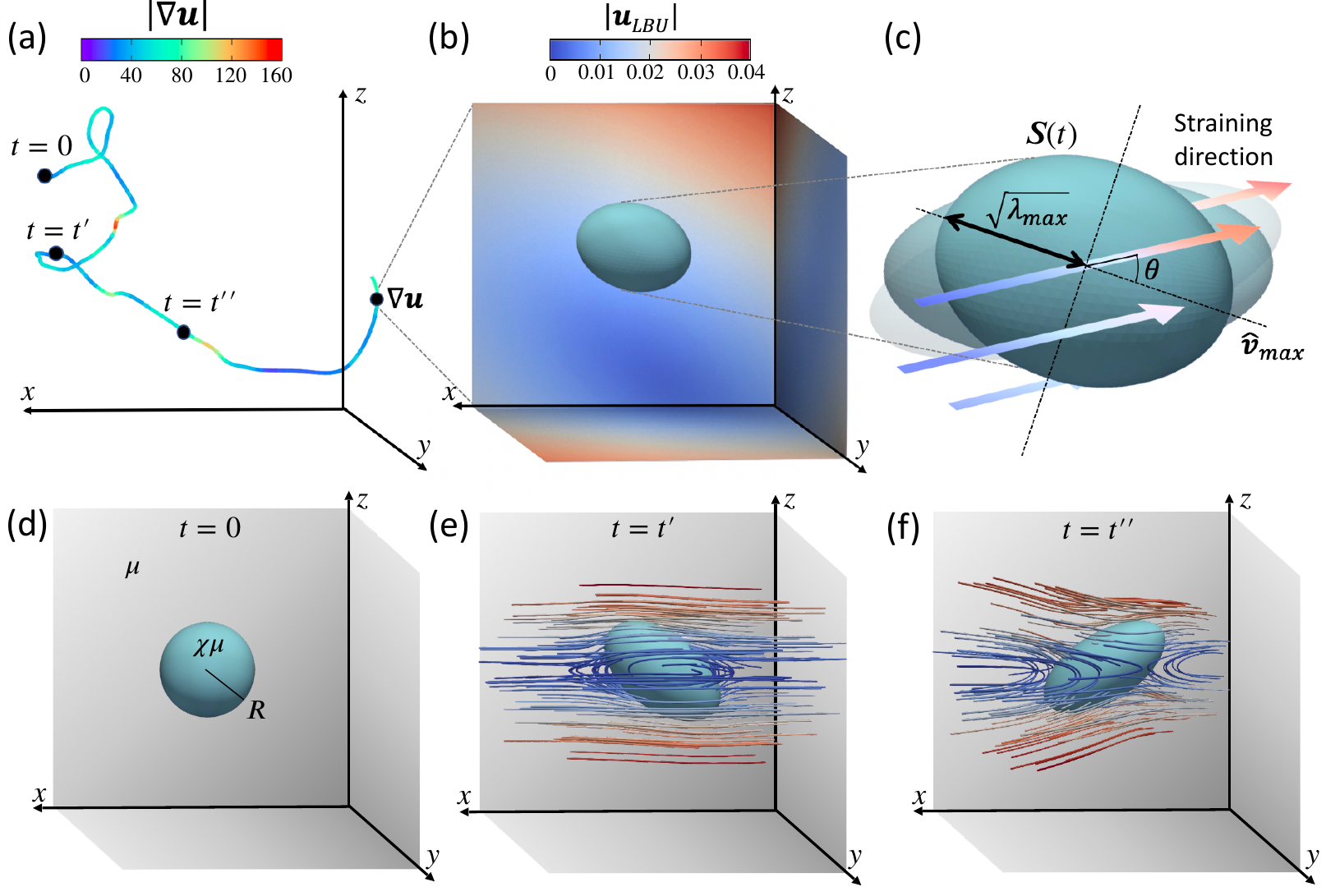}%
\caption{System and numerical setup description: trajectory and velocity gradients $\nabla\vec{u}$ of a droplet are probed from direct numerical simulations (DNS)~\cite{biferale2023turb} (panel (a)). The turbulent velocity gradients are then imposed at the boundaries of the numerical domain used to perform immersed boundary-lattice Boltzmann (IBLB) simulations (panel (b)). Relevant droplet features are indicated in panel (c) (more details are given in the text). Evolution in the IBLB simulation is sketched in panels (d-f) for three different times \corr{$t=0, t=t', t=t''$ corresponding to different turbulent configurations chosen from the represented trajectory in panel (a)}. Flow streamlines are reported with colormap following velocity gradient magnitude (cf. panel (a)). A complete visualization of the evolution of the droplet is shown in the Supplemental Material~\cite{SupplementalMaterial}.}\label{fig:System}%
\end{figure*}

The dynamics of droplets in turbulent flows primarily depends on the ratio between the droplet characteristic scale (e.g. the radius of the droplet at rest, cf. Fig.~\ref{fig:System}) and the Kolmogorov dissipation length $\eta$, expressing the scale at which the energy spectrum of the turbulent flow is cut off~\cite{frisch1995turbulence,sagaut2008homogeneous,Benzi2023}.
Droplets larger than the Kolmogorov length scale are subjected to inertia-dominated hydrodynamic stresses, whereas viscous stresses dominate for droplets smaller than the Kolmogorov scale (sub-Kolmogorov droplets). Here we focus on the latter regime, where the flow deforming the droplet can be described by the Stokes equations matched with the turbulent far field~\cite{cristini2003turbulence,milan2020sub} (cf. Sec.~\ref{sec:assumptions}). This is a crucial assumption for having analytical reference models to compare with, in the limit of small deformations.

The dynamics of sub-Kolmogorov droplets has been investigated with various numerical approaches. The boundary integral approach was used to capture critical sub-Kolmogorov droplet size and breakup rates in~\cite{cristini2003turbulence}, while the lattice Boltzmann (LB) method based on the diffuse-interface multiphase approach was validated for sub-Kolmogorov investigations and employed to model high deformations in Ref.~\cite{milan2020sub}; statistical analysis on droplets dynamics was performed in Refs.~\cite{biferale2014deformation,Ray2018,milan2020sub} based on phenomenological models for ellipsoidal droplet dynamics~\cite{maffettone1998equation}.
In this work, we employ a novel route based on the hybrid coupling between the immersed boundary (IB) method~\cite{peskin2002immersed} and the LB method~\cite{benzi1992lattice,succi2001lattice}. In contrast to multiphase/multicomponent LB approaches~\cite{shan1993lattice,Chen2014}, the hybrid IBLB method could have several advantages: it inherently preserves the sharp-interface limit of hydrodynamics without creating spurious currents~\cite{Connington2012} and it allows to easily introduce various interfacial properties, like the surface viscosity~\cite{li2019finite,guglietta2020effects} or generic constitutive laws~\cite{feng2004immersed,zhang2007immersed,Kruger2011,muller2021hyperelastic,pelusi2023sharp}, while having the possibility of using finer meshes to model them~\cite{Peng2006}. Furthermore, accurate modeling of the outer flow via linearization of turbulent velocity gradients can be easily achieved with the IBLB approach employing elementary LB schemes~\cite{book:kruger}. Overall, if from one side the use of the IBLB method for the simulation of sub-Kolmogorov droplet dynamics in turbulent flows seems appropriate, the applicability of the method in this context has never been quantitatively assessed so far. This paper aims to fill this gap.

The paper is organized as follows: in Sec.~\ref{sec:assumptions}, we review the continuum equations and control parameters of our system; in Sec.~\ref{sec:numerical}, we describe the IBLB method used to simulate the dynamics of a single sub-Kolmogorov droplet in homogeneous isotropic turbulence; simulation results and comparisons with theoretical models in the limit of small deformation are presented in Sec.~\ref{sec:results}; conclusions and a summary of our results will be presented in Sec.~\ref{sec:conclusions}. 

%%%%Fig 1%%%%%%%%%%
%%%%%%%%%%%%%%%%%%%%%%%%%%%%%%%%%%%%%%%%%%%

%%%%%%%%%%%%%%%%%%%%%%%%%%%%%%%%%%%%%%%%%%%%%%%%%%%
\section{\label{sec:assumptions} Problem Statement}
%%%%%%%%%%%%%%%%%%%%%%%%%%%%%%%%%%%%%%%%%%%%%%%%%%%
We examine a system where a droplet with rest radius $\displaystyle R$ and surface tension $\displaystyle \sigma$ is immersed in homogeneous isotropic turbulence (cf. Fig.~\ref{fig:System}). 
The dynamic viscosity of the carrier fluid is $\displaystyle\mu$, while $\displaystyle\chi\mu$ is the dynamic viscosity of the droplet phase, so that $\displaystyle\chi$ denotes the system viscosity ratio. In our analysis, we investigate the deformation of droplets in the centered frame reference moving along a given trajectory. In this description, we assume to be in the sub-Kolmogorov regime, i.e., $R/\eta\ll1$, with $\displaystyle\eta=(\nu^3/\epsilon)^{1/4}$ being the Kolmogorov length scale, $\nu=\mu/\rho$ the kinematic viscosity of the turbulent flow with $\rho$ the fluid density, $\tau_{\eta}=(\nu/\epsilon)^{1/2}$ the Kolmogorov time scale and \corr{$\displaystyle\epsilon=\nu \left\langle \nabla\vec{u}:\nabla\vec{u} \right\rangle$} the turbulent dissipation rate~\cite{frisch1995turbulence,sagaut2008homogeneous,Benzi2023}, with $\displaystyle\langle...\rangle$ indicating the ensemble mean.
\corr{For practical applications, sub-Kolmogorov regime is achieved for droplets smaller than a threshold which can be found via the relation $\eta \sim \Lambda/\Rey^{3/4}$, with $\Lambda$ indicating a characteristic length scale for the considered system. For the applications mentioned in Sec.~\ref{sec:intro}, the typical tanks used for mixing processes and water treatments set $\Lambda$ to meters and $\Rey\sim 10^3$, with ``small droplets'' identified by sub-millimetric drops, whereas in most hemodynamical applications, such as in rotary blood pumps where $\eta\sim 20\times 10^{-6}$ m~\cite{Schle2017}, red blood cells can mostly be considered sub-Kolmogorov ``small droplets''.}

Within the sub-Kolmogorov assumption, the droplet Reynolds number:
\begin{equation}\label{reynolds_drop}
    \Rey_{\mbox{\tiny drop}}=\frac{R^2}{\nu \tau_{\eta}}=\left(\frac{R}{\eta}\right)^{2}\ ,
\end{equation}
is inherently small, indicating the disparity in the inertial stress $\Pi_{\rho}=\rho R^2/\tau_{\eta}^2$ and viscous stress $\Pi_{\mu}=\mu/\tau_{\eta}$, with the latter dominating the overall hydrodynamic stress.
\corr{Given this fact, one can assume 
that the dynamics is governed by the Stokes equation for an incompressible flow:}
\begin{subequations}\label{stokeseq}
\begin{align}
\boldsymbol{\nabla}\cdot\vec{\Pi'}&=\boldsymbol{\nabla}\cdot \vec{u}'=0 &  \mbox{\small inside the droplet,} \\
\boldsymbol{\nabla}\cdot\vec{\Pi}&=\boldsymbol{\nabla}\cdot \vec{u} =0 & \mbox{\small outside the droplet,} 
\end{align}
\end{subequations}
with:
\begin{subequations}\label{stresstensor}
\begin{align}
    &\vec{\Pi}'=-p'\boldsymbol{I}+\chi\mu\left(\boldsymbol\nabla\vec{u}'+\boldsymbol\nabla\vec{u}'^{\dagger}\right)\ , \\
    &\vec{\Pi}=-p\boldsymbol{I}+\mu\left(\boldsymbol\nabla\vec{u}+\boldsymbol\nabla\vec{u}^{\dagger}\right)\ ,
\end{align}
\end{subequations}
being the inner and outer hydrodynamic stress, with $\boldsymbol{I}$ the 3D identity matrix and $\displaystyle p', p$ the inner and outer pressure, respectively. 
Denoting with $\vec{x_c}$ the droplet center, the far-field flow surrounding the droplet matches a linear expansion constructed from the turbulent velocity gradients $\displaystyle\nabla\vec{u}(t)$:
\begin{equation}\label{eq: linearized velocity}
    \vec{u}(\vec{x},t)=\left(\vec{x}-\vec{x_c}\right)\cdot \nabla\vec{u}(t).
\end{equation}
Denoting with $\tau_{\sigma}=\mu R/\sigma$ the droplet characteristic time set by the balance between viscous effects and surface tension, the overall droplet deformation state is controlled by the capillary number:
\begin{equation}\label{capillary}
    \Ca=\frac{\tau_{\sigma}}{\tau_{\eta}}\ .
\end{equation}
For small values of $\Ca$, a small deformation regime is expected, wherein the time scale for \corr{droplet {relaxation towards sphericity}} is much smaller than the characteristic time of the turbulent flow, and the droplet tends to retract very soon to any deformation imposed by the outer flow; on the other hand, for large values of $\Ca$, the droplet can be exposed for longer times to the straining effects of the outer flow and non-linear effects emerge with larger deformations as well as misalignment with the underlying flow~\cite{cristini2003turbulence}.
In our analysis, the state of the droplet is assessed by looking at both its elongation and orientation with respect to the straining direction at all times (cf. Fig. \ref{fig:System}\corr{, where representative times $t=0, t=t', t=t''$ are reported in panels (d-f), respectively}). Information on the deformation is retrieved from the largest eigenvalue $\mathcal{\lambda}_{\mbox{\tiny max}}$ (corresponding to the eigenvector $\vec{\hat{v}}_{\mbox{\tiny max}}$) of the morphological tensor $\boldsymbol{S}$, i.e., the second order tensor describing the droplet shape~\cite{frankel1970constitutive,rallison1980note}:
\begin{equation}
    \boldsymbol{S}\vec{\hat{v}}_{\mbox{\tiny max}}=\lambda_{\mbox{\tiny max}}\vec{\hat{v}}_{\mbox{\tiny max}} \ .
\end{equation} 
Assuming an ellipsoidal shape, the square root of $\mathcal{\lambda}_{\mbox{\tiny max}}$ coincides with the major semi-axis of the droplet (cf. Fig.~\ref{fig:System}). The instantaneous droplet orientation is studied via the orientation parameter $\beta$~\cite{cristini2003turbulence}: 
\begin{equation}\label{beta}
\displaystyle \beta=\frac{\vec{\hat{v}}_{\mbox{\tiny max}}\cdot\vec{E}\cdot\vec{\hat{v}}_{\mbox{\tiny max}}}{||\vec{E}||} \ ,
\end{equation}
where $\vec{E}=\left(\nabla\vec{u}+\nabla\vec{u}^{\dagger}\right)/2$ is the symmetric velocity gradient matrix representing the straining component of the flow, with $||\vec{E}||$ being its Frobenius norm. The orientation parameter $\displaystyle\beta$ weights the effects of the instantaneous strain $\boldsymbol{E}(t)$ on the major elongation direction attained by the droplet at time $t$: $\beta$ depends on the inclination angle $\theta$ between the droplet and the straining direction of the flow (cf. Fig. \ref{fig:System}), it is maximum when $\theta=0$ (complete alignment), and minimum when $\theta=\pi/2$ (complete misalignment). For example, in stationary shear flow $\beta=1/\sqrt{2}$ for a droplet aligned with the strain direction and $\beta=-1/\sqrt{2}$ when the droplet elongation is orthogonal to it. \\
In Sec.~\ref{sec:results}, we compare our numerical results with the linear theory in~\cite{barthes1973deformation,rallison1980note} adapted for turbulent gradients:
\begin{multline}\label{eq:LT}\frac{\partial\vec{S}(t)}{\partial t}=-\frac{1}{\tau_{\sigma}}\frac{40\left(\chi+1\right)}{\left(2\chi+3\right)\left(19\chi+16\right)}\left(\vec{S}(t)-\boldsymbol{I}R^2\right)+\\
+2R^2\frac{5}{2\chi+3}\vec{E}(t)\ .
\end{multline}
This equation is found by assuming an expansion of the morphological tensor $\boldsymbol{S}$ in power series of $\Ca$ plugged into the Stokes equations~\eqref{stokeseq} and retaining only the linear contributions in $\Ca$~\cite{lamb1932hydrodynamics,cox1969deformation}. We will use this equation to get predictions for the droplet deformation and orientation when $\Ca\longrightarrow 0$ and validate the results of IBLB numerical simulations in such a limit.

\section{\label{sec:numerical}Numerical Method}
\subsection{Immersed boundary-lattice Boltzmann (IBLB) model}\label{sec:IBLB+BC}
Various numerical methods have been developed for simulating bulk viscous flows with deformable suspensions, such as boundary element methods~\cite{rallison1978numerical,pozrikidis1992boundary,cristini2003turbulence,art:gounley16}, volume-of-fluid methods~\cite{li2000numerical}, and lattice Boltzmann methods~\cite{book:kruger}. In our work, we employ the immersed boundary-lattice Boltzmann (IBLB) method, which has been extensively used in previous works for investigating viscoelastic capsules~\cite{thesis:kruger,Kruger2011,li2019finite,guglietta2020effects,Guglietta2023,li2021similar,kruger2013crossover,art:kruger14deformability,gekle2016strongly,Silva2024}. The method combines the IB method, used to couple the sharp droplet interface with the fluid, with LB method, which simulates bulk viscous flows. Few works specifically focused on using such a method for simulating droplet dynamics in stationary flows~\cite{li2019finite,guglietta2020effects,taglienti2023reduced,pelusi2023sharp}: here, we take a step further, adapting the method for the simulation of droplet dynamics in a complex turbulent flow.

\begin{figure}[t]
\centering
\includegraphics[width=1.0\linewidth]{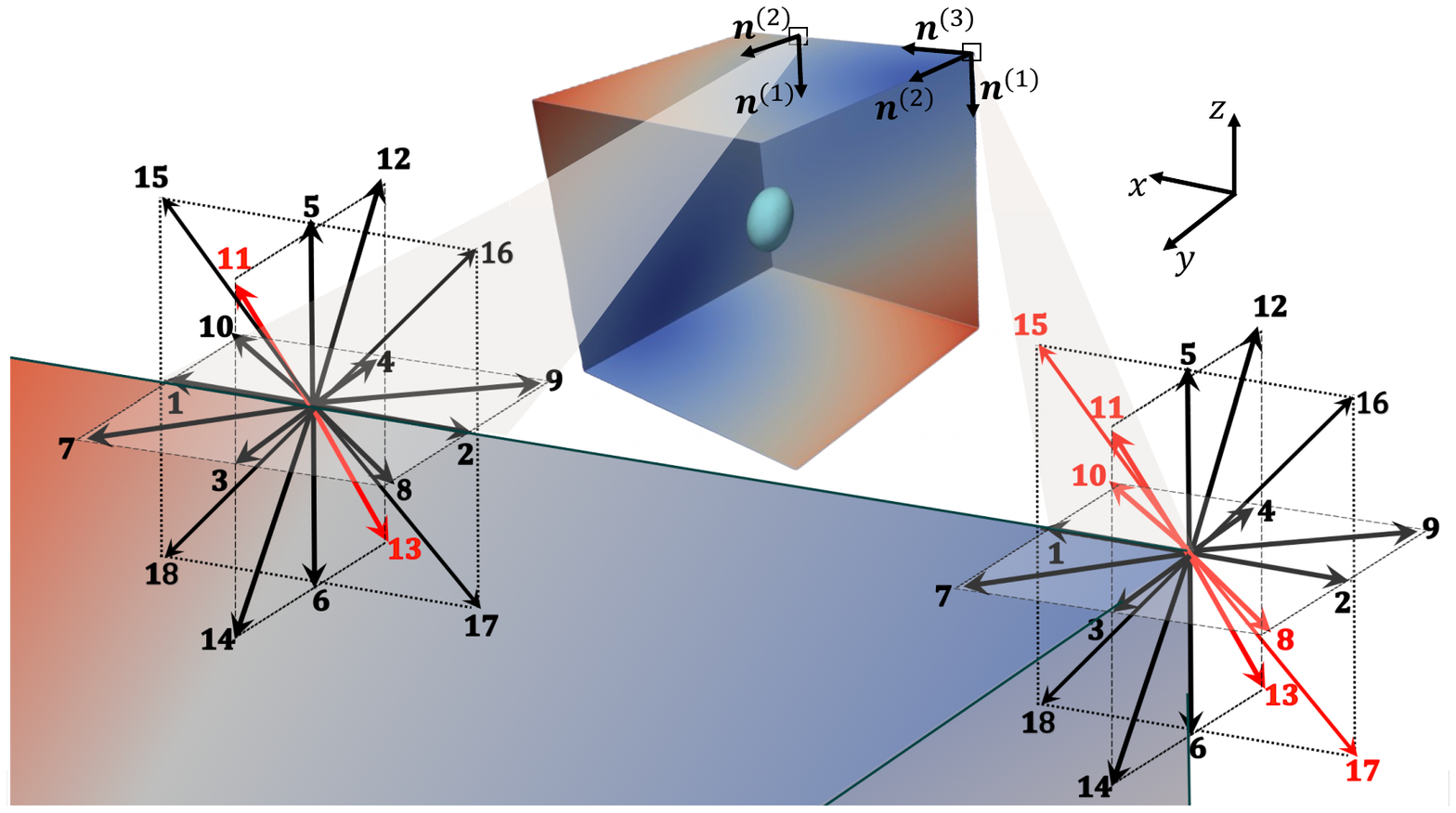}%
\caption{Representation of LB scheme applied for a specific edge (on the left) and corner (on the right). Numbers refer to the LB discrete velocities $\vec{c}_i$. Buried links are highlighted in red. Two examples of vectors $\vec{n}^{(i)}$ normal to the planes are given on the top, for both the edge and the corner.}\label{Fig:ZH sketch}%
\end{figure}

The LB method is a kinetic approach to simulate hydrodynamics, evolving probability distribution functions in discrete directions. In this work, we use the D3Q19 velocity scheme with a set of 19 discrete velocities  $\displaystyle \vec{c}_i$~\cite{book:kruger,succi2001lattice}, which are indicated in Fig.~\ref{Fig:ZH sketch}, with $\vec{c}_0$ related to the resting population. The LB method recovers both the continuity and Navier-Stokes equations when fluctuations around the local equilibrium are small~\cite{book:kruger,succi2001lattice}.
In the following, all quantities are given in lattice Boltzmann units (LBU), in which we assume both the lattice and time spacing to be unitary. In the chosen lattice scheme, the evolution of probability distribution functions $\displaystyle f_i$  with discrete velocities $\vec{c}_i$ at  coordinate $\displaystyle\vec{x}$ and time $\displaystyle t$ is regulated by the LB equation:
\begin{equation}\label{LBMEQ}
f_i(\vec{x}+\vec{c}_i, t+1) - f_i(\vec{x}, t) =\left[\Omega_i(\vec{x},t) + S_i(\vec{x},t)\right]\ ,
\end{equation}
including the streaming along direction $i$ on the l.h.s., the collision term $\Omega_i$ and the force source $S_i$ on the r.h.s. 
Here, we implemented the standard Bhatnagar-Gross-Krook (BGK) collision operator~\cite{qian1992lattice,book:kruger}:
\begin{equation}\label{eq:collision}
\Omega_i(\vec{x},t)=-\frac{1}{\tau_{\mbox{\tiny f}}}[f_i(\vec{x}, t) - f_i^{(\mbox{\tiny eq})}(\vec{x}, t)]\ ,
\end{equation}
stating that distribution functions relax towards equilibrium $f_i^{(\mbox{\tiny eq})}(\vec{x}, t)$ with characteristic time $\tau_{\mbox{\tiny f}}$.
\corr{We remark that a MRT algorithm could also be chosen, as the use of multiple relaxation times can improve the IBLB stability~\cite{Lu2012} and, generally, make it possible to discern the effects of bulk viscosity and compressibility. In our case, these are negligible and left out with a SRT scheme.}
The relaxation time is directly linked to the kinematic viscosity of the system $\nu$~\cite{benzi1992lattice,succi2001lattice}:
\begin{equation}\label{eq:tau_to_nu}
    \nu =  \cs^2\left(\tau_{\mbox{\tiny f}}-\frac{1}{2}\right)\ ,
\end{equation}
where $\cs=1/\sqrt{3}$ is the speed of sound.
The local equilibrium dependency on $\vec{x}$ and $t$ is due to the density $\rho=\rho(\vec{x},t)$ and velocity field $\vec{u}=\vec{u}(\vec{x},t)$~\cite{qian1992lattice}:
\begin{equation}\label{equilibriumLB}
f^{(\mbox{\tiny eq})}(\vec{x},t)= w_i\rho\left(1+\frac{\vec{u}\cdot\vec{c}_i}{c_s^2}+\frac{(\vec{u}\cdot\vec{c}_i)^2}{2c_s^4}-\frac{\vec{u}\cdot\vec{u}}{c_s^2}\right)\ ,
\end{equation}
where $w_i=w(|\vec{c}_i|^2)$ are suitable weights, such that $w_0=1/3$, $w_{1-6}=1/18$, $w_{7-18}=1/36$.
During the relaxation process, external forces $\vec{F}_{\tiny{\mbox{ext}}}=\vec{F}_{\tiny{\mbox{ext}}}(\vec{x}, t)$ can be added in the LB equation via the source term $S_i(\vec{x},t)$ according to the Guo scheme~\cite{guo2002discrete}:
\begin{equation}
S_i(\vec{x},t)=\left(1-\frac{1}{2\tau_{\mbox{\tiny f}}}\right)\frac{w_i}{c_s^2}\left[\left(\frac{\vec{c}_i\cdot \vec{u}}{c_s^2}+1\right)\vec{c}_i-\vec{u}\right]\cdot\vec{F}_{\tiny{\mbox{ext}}}\ .
\end{equation}
The populations $\displaystyle f_i$ can then be used to retrieve the density and momentum fields used to compute the equilibrium and source terms:
\begin{subequations}\label{eq:density_velocity_LB}
\begin{align}\label{eq:density_LB}
&\rho(\vec{x}, t) = \sum_{i} f_i(\vec{x}, t)\; , \\
&\rho\vec{u}(\vec{x}, t) = \sum_{i} \vec{c}_i f_i(\vec{x}, t)+\frac{\vec{F}_{\tiny{\mbox{ext}}}(\vec{x}, t)}{2}\ , \label{eq:velocity_LB}
\end{align}
\end{subequations}
with the half-force correction appearing in the momentum~\cite{book:kruger,guo2002discrete}. \\
The interface of the droplet is represented by a finite set of Lagrangian points forming a 3D triangular mesh with zero thickness. 
On each triangle, the stress $\vec{\Pi}_{\text{\tiny drop}}$ given by the surface tension is:
\begin{equation}\label{eq:pi_drop}
    \vec{\Pi}_{\text{\tiny drop}}=\sigma\boldsymbol{I}_2\ , 
\end{equation}
where $\boldsymbol{I}_2$ is the 2D identity matrix~\cite{li2019finite,guglietta2020effects}.
\corr{Within the representation in Eq.~\eqref{eq:pi_drop}, the tangential forces per unit area coming from surface tension are equally distributed across the droplet surface, naturally preserving the isotropic nature of $\sigma$. The resulting net force arising from tangential forces acting on an infinitesimal surface area points inward with respect to the surface curvature. In the case of a static droplet, this causes the inner pressure to increase homogeneously in the bulk of the droplet, in compliance with the Laplace law.}
Details on the computation of the force $\vec{\varphi}_i$ on the $i-$th Lagrangian point from the stress tensor are given in~\cite{guglietta2020effects}.\\

The two-way coupling between Lagrangian points and fluid nodes is handled via interpolation according to the IB method~\cite{peskin2002immersed,book:kruger}:
\begin{subequations}\label{eq:ibm_force_d}
\begin{align}
&\vec{F}_{\tiny{\mbox{ext}}}(\vec{x},t) =  \sum_i\vec{\varphi}_i(t)\Delta(\vec{r}_i-\vec{x})\ , \\
&\frac{\partial\vec{r}_i(t)}{\partial t} = \sum_{\vec{x}} \vec{u}(\vec{x},t)\Delta(\vec{r}_i-\vec{x})\ ,
\end{align}    
\end{subequations}
where the $\Delta$ function is the discretised Dirac delta and $\vec{r}_i(t)$ represents the coordinates of the $i$-th Lagrangian node. The $\Delta$ function can be factorised in three interpolation stencils $\Delta(\vec{x}) = \phi(x)\phi(y)\phi(z)$ which can be chosen and tweaked for higher accuracy in the two-way coupling; here we use the 4-points interpolation stencil~\corr{\cite{peskin2002immersed}}:
\begin{small}
\begin{equation}
\phi(x) = \begin{cases}
\frac{1}{8}\left(3-2\vert x\vert + \sqrt{1+4\vert x\vert-4x^2} \right) & 0\le \vert x\vert <1\ , \\
\frac{1}{8}\left(5-2\vert x\vert - \sqrt{-7+12\vert x\vert-4x^2} \right) & 1\le \vert x\vert< 2\ , \\
0 & 2\le \vert x\vert\; .
\end{cases}
\end{equation}
\end{small}
 Once the velocity of the Lagrangian points has been computed (cf. Eq.~\eqref{eq:ibm_force_d}), the position $\vec{r}_i(t)$ is updated via an explicit forward Euler method:
\begin{equation}\label{eq:fwd_euler}
    \vec{r}_i(t+1) = \vec{r}_i(t) + \frac{\partial \vec{r}_i(t)}{\partial t}\ .
\end{equation}
%%%%Fig 2%%%%%%%%%%%%%%%%%%%%%%%%%%%%%%%%%%%
%%%%%%%%%%%%%%%%%%%%%%%%%%%%%%%%%%%%%%%%%%
Concerning the boundary conditions, the flux on the boundary domain is fixed at each time $\displaystyle t$ by the turbulent gradients $\nabla\vec{u}(t)$ (cf. Eq.~\eqref{eq: linearized velocity}): the linearization of the velocity fields is centered within the droplet center of mass (cf. Fig.~\ref{fig:System}).
In particular, we set:
\begin{small}
\begin{subequations}\label{eq:velo_facce_cubo_da_gradiente}
\begin{align}
&u_x=\frac{\partial u_x}{\partial x}\left(l_x-\frac{L_x}{2}\right)+\frac{\partial u_x}{\partial y}\left(l_y-\frac{L_y}{2}\right)+\frac{\partial u_x}{\partial z}\left(l_z-\frac{L_z}{2}\right) \ , \\
&u_y=\frac{\partial u_y}{\partial x}\left(l_x-\frac{L_x}{2}\right)+\frac{\partial u_y}{\partial y}\left(l_y-\frac{L_y}{2}\right)+\frac{\partial u_y}{\partial z}\left(l_z-\frac{L_z}{2}\right) \ , \\
&u_z=\frac{\partial u_z}{\partial x}\left(l_x-\frac{L_x}{2}\right)+\frac{\partial u_z}{\partial y}\left(l_y-\frac{L_y}{2}\right)+\frac{\partial u_z}{\partial z}\left(l_z-\frac{L_z}{2}\right) \ ,
\end{align}
\end{subequations}
\end{small}
where $L_x, L_y$ and $L_z$ represent the side lengths of the computational domain, and $l_i\in[0,L_i]$ with $i=x,y,z$. Each boundary plane is uniquely determined by fixing one of the three indices $l_i$ at a value $0$ or $L_i$, while varying the other two $l_j$ in the range $\left[0,L_j\right]$ (with $j\ne i$). This is required to impose a matching of the system with the turbulent far-field, and the velocity in the bulk domain is then naturally developed via the LB dynamics. When performing the streaming step in the LB method (cf. Eq.~\eqref{LBMEQ}) on the nodes at the boundary, special treatment is required for those populations pointing out of the domain, whereas those in the opposite direction are somewhat undefined as there is no lattice node where they would come from. To implement boundary conditions for the outer flow, we adopt the Zou-He velocity boundary conditions~\cite{Zou1997}. In particular, since the original proposed scheme is for the D2Q9 scheme, we employ the 3D generalization for generic flow directions presented in Ref.~\cite{Hecht2010}. 
The implementation of such boundary conditions allows to impose generic time dependent flows at the system boundaries regardless of their instantaneous straining direction:
this is achieved by supporting a regular bounce back rule with a counter slip in the reflected populations at the boundary nodes, acting as a transversal momentum correction. 
Among all the populations at the boundary, we distinguish between three different cases: (a) populations that are neither in the corners nor in the edges, (b) populations that are in the edges but not in the corner, and (c) populations that are in the corners.
Let us introduce the tangential vectors $\displaystyle\vec{t}_i=\vec{c}_i-\left(\vec{c}_i\cdot \vec{n}\right)\vec{n}$, where $\vec{n}$ represents the normal vector to a boundary plane.
We indicate with $f_i$ populations corresponding to the $i$-th discrete velocity $\vec{c}_i$ streaming outside of the domain, and with $f_{-i}$ those that get reflected inside the domain, whose discrete velocity is $-\vec{c}_i$.
For case (a), the reflected populations $\displaystyle f_{-i}$ read~\cite{Hecht2010}:
\begin{multline}\label{ZH_bulk}
    f_{-i}=f_i-\left(2-|\vec{t}_i|\right)\frac{\rho}{6}\left(\vec{c}_i\cdot\vec{u}\right)-\frac{\rho}{3}\left(\vec{t}_i\cdot\vec{u}\right)+\\
    +\frac{1}{2}\sum_{j=0}^{18}f_j\left(\vec{t}_i\cdot\vec{c}_j\right)\left(1-|\vec{c}_j\cdot\vec{n}|\right) \ .
\end{multline}
As can be seen for those populations pointing perpendicular w.r.t. the boundary plane (i.e., $\vec{t}_i=\vec{0}$), we remark that the factor $2-\vert\vec{t}_i\vert$ must be included in the second term of the r.h.s. to retrieve the standard bounce-back rule~\cite{book:kruger}:
\begin{equation}\label{standard BB}
    f_{-i}=f_i-\frac{\rho}{3}\left(\vec{c}_i\cdot\vec{u}\right) \ .
\end{equation}
For the case (b), two different planes with normal vectors $\vec{n}^{(1)}$ and $\vec{n}^{(2)}$ must be taken in consideration (cf. Fig.~\ref{Fig:ZH sketch}), with the reflected populations that read:
\begin{multline}\label{ZH_edges}
f_{-i}=f_i-\frac{1}{4}\sum_{j=0}^{18}f_j\left(\vec{t}_i\cdot\vec{c}_j\right)\\
\cdot \left(1-|\vec{c}_j\cdot\vec{n}^{(1)}|\right)\left(1-|\vec{c}_j\cdot\vec{n}^{(2)}|\right)\ .
\end{multline}
Among these populations, we define $f^{(1,2)}$ those associated with the two lattice vectors $\vec{c}^{(1,2)}=\pm\left(\vec{n}^{(1)}-\vec{n}^{(2)}\right)$. The latter constitute the so-called ``buried links'', i.e., those lattice vectors along which both $f_i$ and $f_{-i}$ points outside the domain (see red arrows in Fig.~\ref{Fig:ZH sketch}). For such populations, Eq.~\eqref{ZH_edges} does not hold, and an additional prescription is required ~\cite{Hecht2010}:
\begin{multline}\label{zh-buried-links-edges}
    f^{(1,2)}=\frac{1}{22}\sum_{i=1}^{18}f_i\Bigg\{1-\left(1-\left|\vec{c}_i\cdot\left[\vec{n}^{(1)}\times\vec{n}^{(2)}\right]\right|\right)\\
    \cdot \left(1-\left|\vec{c}_i\cdot\left[\frac{\vec{n}^{(1)} +\vec{n}^{(2)}}{|\vec{c}_i|^2}\right]\right|\right)\Bigg\}  \ .
\end{multline}
On each edge, there is exactly one population $f^{(1)}$ streaming along $\vec{c}^{(1)}$ with the corresponding opposite one $f^{(2)}$ streaming along $\vec{c}^{(2)}$. 
As an example, in Fig.~\ref{Fig:ZH sketch}, we consider the edge in $y=0$ and $z=L_z$: the two normals are $\vec{n}^{(1)}=(0,0,-1)$ and $\vec{n}^{(2)}=(0,1,0)$. The buried links are therefore identified by $\vec{c}^{(1,2)}=\pm(0,-1,-1)$, that correspond to the directions $\vec{c}_{11}$ and $\vec{c}_{13}$ (red arrows in Fig.~\ref{Fig:ZH sketch}).
The resting population $f_0$ must be corrected as $\displaystyle f_0=12f^{(1,2)}$~\cite{Hecht2010}.
Lastly, we consider case (c). For each corner, we have three different planes with normals $\displaystyle \vec{n}^{(1)}, \vec{n}^{(2)}$ and $\vec{n}^{(3)}$ which identify 6 buried links (cf. Fig.~\ref{Fig:ZH sketch}) whose corresponding lattice vectors are $\vec{c}^{(1,\dots,6)}$. The buried vectors are such that $\vec{c}^{(i)}\cdot (\vec{n}^{(1)}+ \vec{n}^{(2)}+ \vec{n}^{(3)})=0$. The corresponding distributions assigned to these six buried vectors are:\begin{equation}\label{zh-buried-links-corners}
    f^{(1,\dots,6)}=\frac{1}{18}\sum_{i=1}^{18}f_i\left|\vec{c}_i\cdot\frac{\vec{n}^{(1)}+\vec{n}^{(2)}+\vec{n}^{(3)}}{|\vec{c}_i|^2}\right| \ .
\end{equation} 
Again, the resting population needs a special consideration, and it is set to $f_0 = 12f^{(1,\dots,6)}$~\cite{Hecht2010}.

\corr{We remark that other implementations of boundary conditions to accommodate the outer flow would be possible. For example, in Ref.~\cite{milan2020sub}, a diffuse interface LB approach was used to investigate droplet deformation in turbulent flows, mainly focusing on deviations from ellipsoidality~\cite{maffettone1998equation}. The boundary condition used in Ref.~\cite{milan2020sub} to accommodate the outer flow comprises the use of ghost boundaries to set the LB populations at equilibrium, serving as a ``thermalization'' of the domain to the outer turbulent field. Preliminary analysis on statistics of droplet deformation revealed that this kind of boundary condition does not guarantee a precise reconstruction of theoretical predictions for small $\Ca$, hence we decided to implement a Zou-He velocity boundary condition~\cite{Zou1997}.}

%%%%%%%%%%%%%%%%%%%%%%%%%%%%%%%%%%%%%%%%%%%%%%%%%%%%%%%%%%%%%%%%%%%%%%%%%%%%%%%%%
\subsection{Outer Flow and matching between DNSU and LBU}\label{subsec:OuterFlow}
%%%%%%%%%%%%%%%%%%%%%%%%%%%%%%%%%%%%%%%%%%%%%%%%%%%%%%%%%%%%%%%%%%%%%%%%%%%%%%%%%
We consider the dynamics of sub-Kolmogorov droplets with $\displaystyle\chi=1$ in an ensemble of turbulent trajectories taken from the TURB-Lagr open-source database~\cite{biferale2023turb}. This database contains over $3\times 10^5$ particle trajectories obtained for the case of homogeneous isotropic turbulence via direct numerical simulations (DNS) of the incompressible Navier-Stokes equations using a pseudo-spectral approach~\cite{Buzzicotti2016}. The Reynolds number calculated based on the Taylor scale is $\displaystyle\Rey\sim 130$. In what we refer to as direct numerical simulation units ($\displaystyle \mbox{DNSU}$), the Kolmogorov length scale is $\eta=4.2\times 10^{-3} \ \mbox{DNSU}$, and the Kolmogorov time scale is $\displaystyle \tau_{\eta}=2.3\times 10^{-2} \ \mbox{DNSU}$. The kinematic viscosity of the system is $\displaystyle \nu=8\times 10^{-4}\ \mbox{DNSU}$, and the dissipation rate is $\epsilon=1.4 \ \mbox{DNSU}$. Each trajectory lasts for roughly an eddy turnover time $\displaystyle \tau_{\mbox{\tiny eddy}}=4.5 \ \mbox{DNSU}\sim 200\tau_{\eta}$ so that, by subjecting droplets to a linearization of the turbulent field as in Eq.~\eqref{eq: linearized velocity}, all the relevant turbulent features are probed along each trajectory.
The local velocity gradients are probed every $\displaystyle \Delta t=2.25\times 10^{-3} \ \mbox{DNSU}$ for a total of $\displaystyle 2\times 10^3$ turbulent realizations for each trajectory.

In order to match the turbulent outer flow with the IBLB simulation, we need to apply a unit conversion from the quantities of the DNS to those used in the IBLB simulations.
Hereafter, a generic quantity $A$ is assumed to be in DNSU, whereas the corresponding quantity in LB units (LBU) is represented as ${A}_{\text{\tiny{LBU}}}$. Therefore, the conversion factor $C_A$ in DNSU is such that $A = A_{\text{\tiny{LBU}}}\ C_A$.
We start by fixing the conversion factor for the length $C_x$.
By fixing a desired value of $R/\eta$, and knowing the Kolmogorov lengthscale $\eta$, the radius of the droplet is known in DNSU and the conversion factor $C_x$ can be computed as:
\begin{equation}\label{eq:cl}
C_x =\frac{R}{R_{\text{\tiny{LBU}}}}\ .
\end{equation}
Then, we consider the unit conversion for time, which can be retrieved from the viscosity. Indeed, since $\nu =\nu_{\text{\tiny{LBU}}} C_x^2/C_t$, we have: 
\begin{equation}\label{eq:ct}
    C_t = C_x^2\frac{\nu_{\text{\tiny{LBU}}}}{\nu}\ .
\end{equation}
Within this prescription, at every time frame $\Delta t$, we read the velocity gradient in DNSU and convert it in LBU as:
\begin{equation}\label{eq:strain_conversion}
\left(\nabla\vec{u}\right)_{\text{\tiny{LBU}}} = C_t\nabla\vec{u}\ .
\end{equation}
For each realization of the velocity gradient, we then perform $N$ IBLB time steps given by: 
\begin{equation}
    N = \frac{\Delta t}{C_t}\ .
\end{equation}
Note that, by using Eq.~\eqref{eq:cl} and Eq.~\eqref{eq:ct}, we obtain that the number of IBLB steps needed to match the dynamics of the outer flow scales with the inverse square of $R/\eta$: 
\begin{equation}\label{eq:N_Reta}
    N = \Delta t\frac{\nu}{\nu_{\text{\tiny{LBU}}}}\left(\frac{R_{\text{\tiny{LBU}}}}{ R}\right)^2 \sim \left(\frac{\eta}{R}\right)^2 \ .
\end{equation}
For a given realization of $\displaystyle \nabla\vec{u}(t)$, convergence to the Stokes dynamics given by Eq.~\eqref{stokeseq} is expected when $R/\eta \rightarrow 0$. As discussed in Sec.~\ref{sec:results}, a good convergence is already observed when $R/\eta \approx 1/2$, corresponding to $N \approx 500$. We remark that by increasing the LB relaxation time $\displaystyle\tau_{\mbox{\tiny f}}$ (and consequently $\displaystyle\nu_{\mbox{\tiny LBU}}$), the number of steps $N$ required for LB to reproduce sub-Kolmogorov droplets dynamics can be reduced without altering the simulation results, albeit attention should be used to not induce over relaxation~\cite{book:kruger}.\\
Once the unit conversion between the two systems has been set, one only needs to compute the surface tension of the drop $\sigma$ in LBU. 
Since we know the value of $\tau_{\eta}$ in DNSU, we can compute $\tau_{\sigma}$ in DNSU for a given capillary number by using its definition given in Eq.~\eqref{capillary}, and convert it in LBU:
\begin{equation}
    \tau_{\sigma,\text{\tiny{LBU}}} = \frac{\tau_{\sigma}}{C_t}\ . 
\end{equation}
The value of the surface tension in LBU is then computed as: 
\begin{equation}\label{surfacetensionconversion}
    \sigma_{\text{\tiny{LBU}}} = \frac{\mu_\text{\tiny{LBU}}R\text{\tiny{LBU}}}{\tau_{\sigma,\text{\tiny{LBU}}}}\ . \\
\end{equation}
Once the parameters $R/\eta, \Ca, \tau_{\mbox{\tiny f}}$, $R_\text{\tiny{LBU}}$ and the computational domain size $L_x\times L_y\times L_z$ are set, one needs to apply the following steps:
\begin{enumerate}
    \item Consider the turbulent gradient matrix ${\nabla\vec{u}}(t)$ at time $t$; 
    \item Apply the conversion in Eq.~\eqref{eq:strain_conversion} to initialize the boundary planes via Eq.~\eqref{eq:velo_facce_cubo_da_gradiente}; 
    \item Perform $N$ IBLB steps with the prescriptions in Sec.~\ref{sec:numerical} employing the surface tension given in Eq.~\eqref{surfacetensionconversion}; 
    \item Consider the turbulent gradient matrix ${\nabla\vec{u}}(t+\Delta t)$ at time $t+\Delta t$ and re-iterate steps 2-4.
\end{enumerate}
In our IBLB setup, we employed a 3D box with size $\displaystyle L_x=L_y=L_z=200\ \mbox{LBU}$ for simulating droplets with rest radius $\displaystyle R_{\text{\tiny{LBU}}}=19\ \mbox{LBU}$ possessing a structured mesh composed of 20480 triangles: these numerical parameters were chosen after performing LB simulations to test the convergence of the LB to the hydrodynamic equations~\cite{succi2001lattice,book:kruger}.
Furthermore, the chosen parameters need to be tailored to minimize confinement effects~\cite{shapiraLowReynoldsNumber1990,Kruger2011} and discrete curvature effects~\cite{li2019finite,Kruger2011}, and a very refined mesh is required for the latter. The value for the LB relaxation time is $\tau_{\mbox{\tiny f}}=1.5$ LBU. 

\corr{Numerical simulations are performed with a GPU in-house code largely employed and benchmarked in previous works~\cite{guglietta2020effects,Guglietta2021,guglietta2021loading,taglienti2023reduced}.}
Simulations ran on Nvidia Ampere ``A100'' Graphic Processing Units (GPUs), with a simulation time for each trajectory lasting roughly 3 hours. 
%%%%%%%%%%%%%%%%%%%%%%%%%%%%%%%%%%%
\begin{figure}[t!]
\centering
\includegraphics[width=1.0\linewidth]{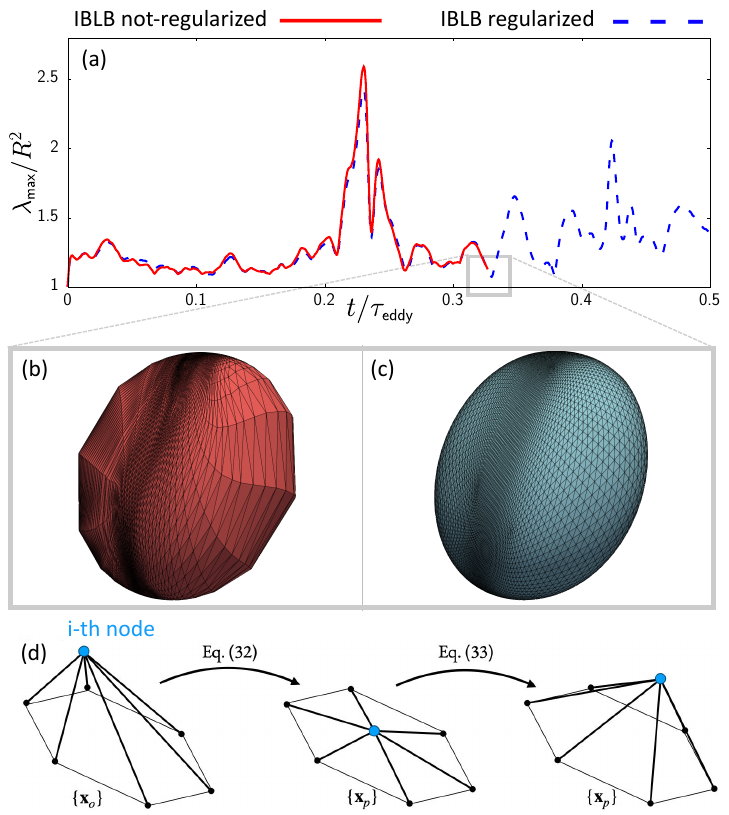}
\caption{Sketch of the mesh regularization employed. Panel (a) shows the same trajectory for a not-regularized (solid line) and regularized droplet (dashed line); panels (b) and (c) represent the Lagrangian mesh right before the time where the not-regularized simulation destabilizes. Panel (d) sketches the regularization process for a specific node and its neighbours (detailed equations are given in the text).}\label{Fig:regularization}
\end{figure}
%%%%%%%%%%%%%%%%%%%%%%%%%%%%%%%%
\subsection{Mesh regularization}\label{subsec: regularization}
%%%%%%%%%%%%%%%%%%%%%%%%%%%%%%%%
The IBLB method, while inherently preserving the sharp-interface limit of hydrodynamics, presents challenges in tracking the evolution of mesh elements, especially in the presence of irregular distribution of surface nodes~\cite{Zinchenko1997,Schmidt2002}.
As shown in~\cite{huaEffectConfinementDroplet2013}, droplets subjected to a prolonged shearing strain rate can incur an accumulation of mesh nodes and very loose mesh elements stretched in the major strain direction. Both issues must be avoided when using the IBLB method to couple fluid and interface: indeed, an extreme clustering of the mesh nodes can cause them to overlap, whereas triangles that get too loose can compromise the two-way coupling with the inner/outer fluid penetrating the mesh.
In Fig.~\ref{Fig:regularization}, panel (a), we report the largest normalized eigenvalue $\displaystyle \lambda_{\mbox{\tiny max}}/R^2$ as a function of the normalized time $t/\tau_{\mbox{\tiny eddy}}$. The solid red line corresponds to an IBLB simulation without mesh regularization, while the dashed blue line represents one that has been regularized using the algorithm described in this section (details provided below). In panel (b), we report the triangular mesh of the drop without regularization right before the simulation became unstable, which shows both loosening and clustering of triangles. In panel (c), the regularized mesh is reported.
Concerning the algorithms employed to obtain regular meshes, one can ideally split them into two big families: the first one involves changes in the mesh topology (e.g., creation/destruction of nodes/triangles); the second one employees smoothing techniques, i.e., without any change in the mesh topology, its vertices are moved in order to obtain a regular mesh. 
In the specific context of IBLB simulations of drops, a hybrid version has been implemented in Ref.~\cite{huaEffectConfinementDroplet2013}: by using an iterative algorithm, the mesh nodes are moved until reaching a given tolerance; then, a ``flipping algorithm'' causing a change in the connections of the mesh nodes is applied. We note that topology changes may generate quite delicate issues since the topology of the drop at rest is essential for the computation of the nodal forces starting from the stress on each triangle (cf. Eq.~\eqref{eq:pi_drop})~\cite{guglietta2020effects,li2019finite,art:yazdanibagchi13}. Hence, we preferred not to change the topology but instead use the extension of the Laplacian smoothing technique proposed by Ref.~\cite{Vollmer1999}. This technique acts globally on each node with an iterative process: after having updated all the mesh nodes via the IBLB step (cf. Eq.~\eqref{eq:fwd_euler}), we end up with a set of mesh coordinates $\{\vec{x}_o\}$. 
Before starting the Laplacian smoothing, we initialize a set of coordinates $\{\vec{x}_p\}=\{\vec{x}_o\}$.  We then apply the following iterative algorithm:
\begin{enumerate}
    \item we initialise the auxiliary set of coordinates $\{\vec{x}_q\}=\{\vec{x}_p\}$. This set is necessary to perform a simultaneous update and represents the mesh coordinates at the previous step of this iterative algorithm. 
    \item We update each node in the set $\{\vec{x}_p\}$:
    \begin{equation}\label{extended laplacian smoothing}
    \vec{x}_p^i=\alpha\vec{x}_o^i+\frac{1-\alpha}{N_i}\sum_{j\in{\tiny \mbox{neigh}}(i)} \vec{x}_q^{j} \ ,
    \end{equation}
    with $\displaystyle \alpha\in [0,1]$ being a user-defined parameter regulating the influence of the starting position of the node~\cite{Vollmer1999}. The sum runs over the $N_i$ neighbors of the $i$-th node. This step represents the classic Laplacian smoothing~\cite{Vollmer1999}.
    \item Once all the coordinates $\vec{x}_p^i$ have been updated, one ends up with some nodes that might have been moved inward or outward w.r.t. the coordinates at the previous iteration $\vec{x}_q^i$ (cf. Fig.~\ref{Fig:regularization}, panel (d)). This may cause the creation of artificial valleys and ridges, thus undermining the local curvature of the droplet and the overall volume conservation~\cite{Vollmer1999}. 
    In order to preserve curvature regularity, we employ the correction to the classic Laplacian smoothing introduced in Ref.~\cite{Vollmer1999} (cf. Fig.~\ref{Fig:regularization}, panel (d)). We therefore correct the coordinates $\vec{x}_p^i$ in the following way:
    \begin{equation}\label{HC_algo_full}
    \qquad\qquad \vec{x}_p^i\mapsto\vec{x}_p^i-\left(\gamma\vec{x}_b^i-\frac{1-\gamma}{N_i}\sum_{j\in{\tiny \mbox{neigh}}(i)}\vec{x}_b^{j}\right) \ ,
    \end{equation}
    where 
    \begin{equation}\label{distanceregularization}
    {\vec{x}_b^i}=\vec{x}_p^{i}-\left[\alpha\vec{x}_o^{i}+(1-\alpha)\vec{x}_q^{i}\right] \ .
    \end{equation}
    The parameter $\displaystyle \gamma\in [0,1]$ is an additional control parameter.
    \item We finally need a condition to exit the iterative algorithm. For each node $\vec{x}_p^i$ in the set  $\{\vec{x}_p\}$, we compute the difference w.r.t the set of coordinates at the previous iteration $\{\vec{x}_q\}$. We exit the loop if the $\max_i|\vec{x}_p^i-\vec{x}_q^i|$ is less than a certain threshold. Otherwise, we repeat the present algorithm from step 1.
\end{enumerate}
In our simulations, we find $\displaystyle \alpha=0.1, \gamma=0.51$ to be a suitable set of parameters, although other options can be taken into consideration~\cite{Bacciaglia2021}. Concerning the condition to break the iterative algorithm, the threshold has been set to 0.01 LBU.
%%%%Fig 4%%%%%%%%%%%%%%%%%%%%%%%%%%%%%%%%%%%
\begin{figure}[t]
\centering
\includegraphics[width=1.0\linewidth]{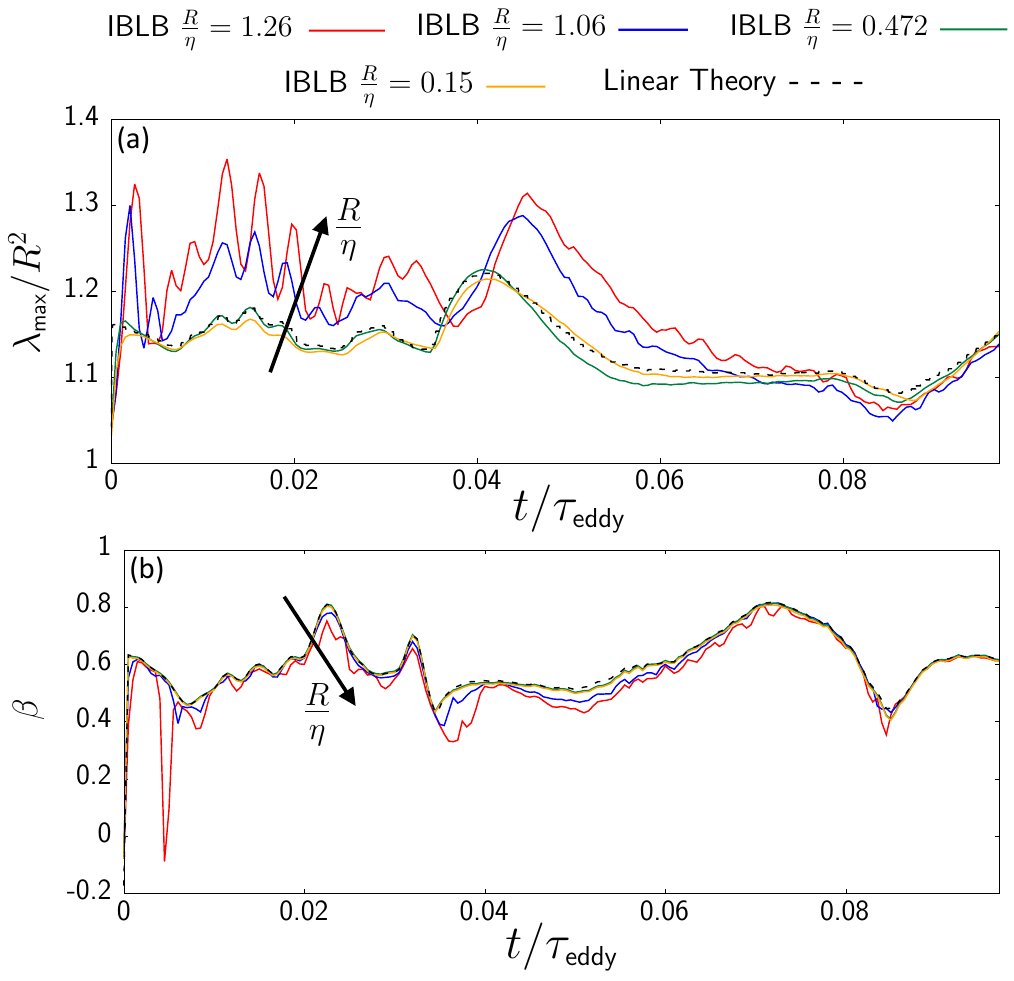}%
\caption{Largest normalized eigenvalue $\displaystyle \lambda_{\mbox{\tiny max}}/R^2$ (panel (a)) and orientation parameter $\displaystyle\beta$ (panel (b)) for a representative trajectory as a function of time for $\Ca=0.05$ and different values of $R/\eta$. Time is made dimensionless with respect to the eddy turnover time $\displaystyle \tau_{\mbox{\tiny eddy}}$ of the outer turbulent flow. Linear theory results obtained from the Stokes equations (cf. Eq.~\eqref{eq:LT}) are also reported. Arrows indicate an increase in $R/\eta$\corr{, with the upper line corresponding to the maximum $R/\eta$ reported in panel (a); hierarchy is reversed in panel (b).}}\label{Fig:singletraj_convergence}
\end{figure}
%%%%%%%%%%%%%%%%%%%%%%%%%%%%%%%%%%%%%%%%%%%

%%%%%%%%%%%%%%%%%%%%%%%%%%%%%%%%%%%%
\section{Results}\label{sec:results}
%%%%%%%%%%%%%%%%%%%%%%%%%%%%%%%%%%%%
We start our numerical investigations by verifying the convergence towards the Stokes equations~\eqref{stokeseq} at changing the relative importance of the droplet radius $R$ with respect to the Kolmogorov scale $\eta$. To this aim, we perform numerical simulations for a fixed $\Ca=0.05$ at changing the ratio $R/\eta$. 
Results for the largest eigenvalue $\displaystyle \lambda_{\mbox{\tiny max}}$ (normalized with $\displaystyle R^2$) and the orientation parameter $\displaystyle\beta$ for a representative trajectory are reported in Fig.~\ref{Fig:singletraj_convergence}. 
The value of $\Ca$ is chosen small enough to ensure that the linear theory (cf. Eq.~\eqref{eq:LT}) holds when $R/\eta \rightarrow 0$, making it possible to have \corr{theoretical} predictions to compare with the IBLB numerical simulations.
Results for linear theory are obtained by integrating Eq.~\eqref{eq:LT} with a standard explicit Runge-Kutta method~\cite{Butcher1996}. By decreasing $R/\eta$, a very good convergence of the IBLB numerical simulations to the linear theory predictions is observed. In the simulated time range, $\displaystyle \lambda_{\mbox{\tiny max}}$ is shown to be more sensible to variations in $\displaystyle R/\eta$, with an oscillating behavior emerging when increasing $\displaystyle R/\eta$ around unity. 
These oscillations emerge alongside both enhanced and delayed elongations, as well as faster retractions, indicating an inability to adapt to the linearized Stokes solutions when $R/\eta \ge 1$.
Milder oscillations are shown for the orientation parameter $\beta$, where departure from the linear theory is less evident at increasing $\displaystyle R/\eta$. By further decreasing $\displaystyle R/\eta$ the computational cost increases (cf. Eq.~\eqref{eq:N_Reta}) but the convergence to the linear theory results does not improve sensibly (see the comparison between the reported case $\displaystyle R/\eta=0.472$ and $\displaystyle R/\eta=0.15$ in Fig.~\ref{Fig:singletraj_convergence}). 

%%%%Fig 5%%%%%%%%%%%%%%%%%%%%%%%%%%%%%%%%%%%
\begin{figure}[th!]
\centering
\includegraphics[width=1.\linewidth]{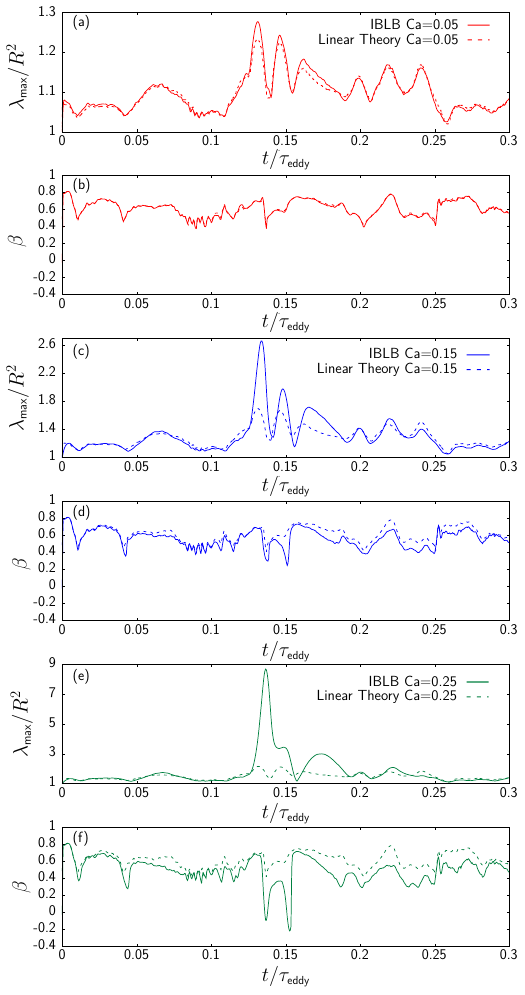}%
\caption{Largest normalized eigenvalue $\displaystyle \lambda_{\mbox{\tiny max}}/R^2$ and orientation parameter $\displaystyle\beta$ for a representative trajectory as a function of time for $R/\eta=0.472$ at changing $\Ca$: $\Ca=0.05$ (panels (a-b)), $\Ca=0.15$ (panels (c-d)) and $\Ca=0.25$ (panels (e-f)). Time is made dimensionless with respect to the eddy turnover time $\displaystyle \tau_{\mbox{\tiny eddy}}$ of the outer turbulent flow. Linear theory results obtained from the Stokes equations (cf. Eq.~\eqref{eq:LT}) are also reported. For a complete visualization of the IBLB simulation for the higher $\Ca$ reported, see Supplemental Material~\cite{SupplementalMaterial}.}\label{Fig:singletraj_results}
\end{figure}
%%%%%%%%%%%%%%%%%%%%%%%%%%%%%%%%%%%%%%%%%%%

%%%%Fig 6%%%%%%%%%%%%%%%%%%%%%%%%%%%%%%%%%%%
\begin{figure*}[th!]
\centering
\includegraphics[width=1.0\linewidth]{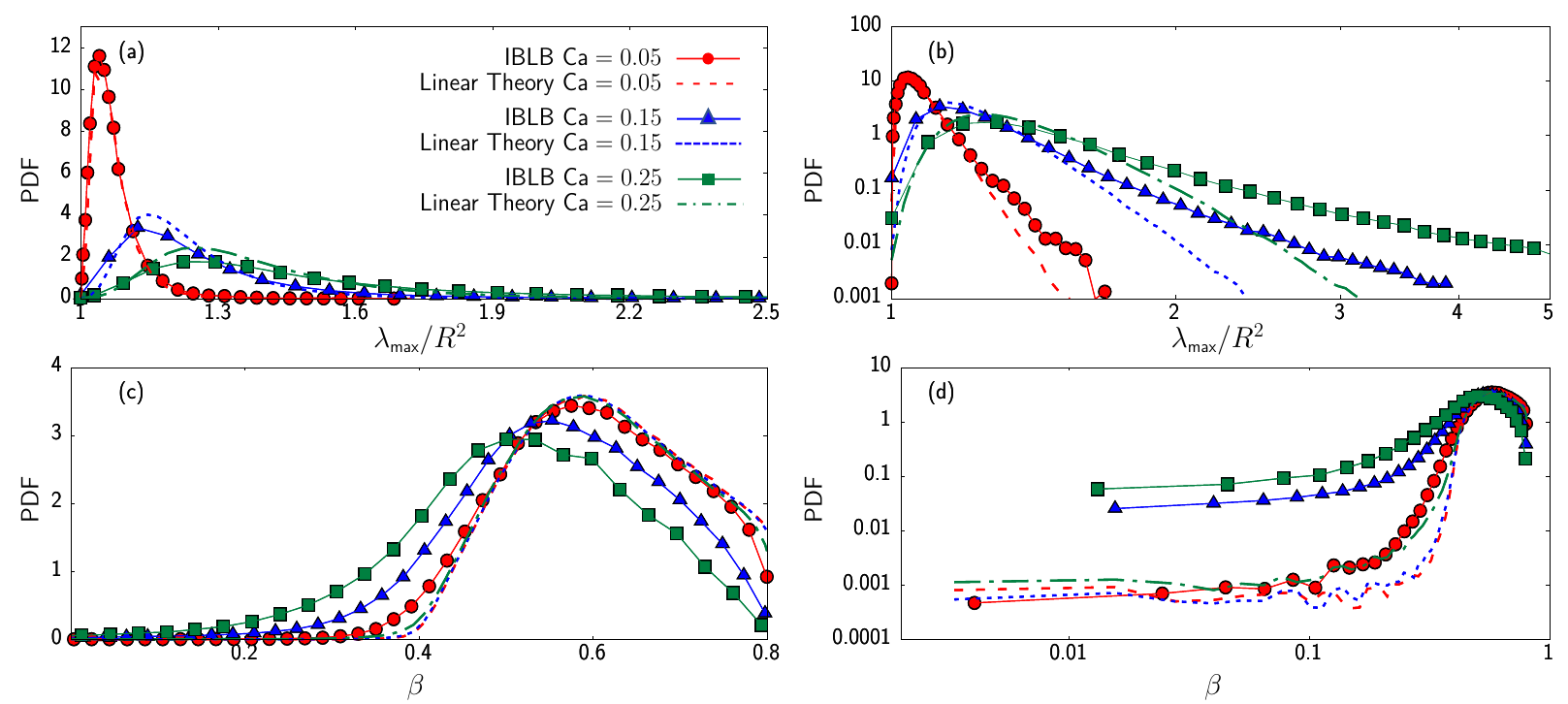}%
\caption{PDF (probability density function) for the largest normalized eigenvalue $\lambda_{\mbox{\tiny max}}/R^2$ and the orientation parameter $\displaystyle\beta$, with $R/\eta=0.472$ and $\Ca=0.05,0.15,0.25$. Both lin-lin plots (panels (a, c)) and log-log plots (panels (b, d)) are reported.}\label{Fig:PDF_results}
\end{figure*}
%%%%%%%%%%%%%%%%%%%%%%%%%%%%%%%%%%%%%%%%%%%
In Fig.~\ref{Fig:singletraj_results}, we complement the findings reported in Fig.~\ref{Fig:singletraj_convergence} by comparing results of IBLB numerical simulations and the predictions of linear theory for $R/\eta=0.472$ at changing $\Ca$. For growing values of $\Ca$, agreement with the linear theory valid for $\Ca\ll 1$ is not achieved, as expected. The value of $\displaystyle \lambda_{\mbox{\tiny max}}$ obtained from IBLB exceeds the linear theory with enhanced peaks. Notice that a similar behavior is found when studying the stationary deformation at finite $\Ca$~\cite{torza1972particle,guido1998three,jackson2003model,li2000numerical,art:gounley16,taglienti2023reduced}.
Large variations in $\displaystyle \lambda_{\mbox{\tiny max}}$ observed in panels (c) and (e) are clearly related to the intermittency in the strain rates, and they become more evident and more persistent in time at increasing $\Ca$. Good agreement in $\beta$ is shown between IBLB numerical simulations and predictions of linear theory up to $\Ca=0.15$, with an overall positive value of $\beta$, pointing to the alignment of the droplet with the straining direction of the outer flow. When $\Ca$ is increasing,  persistence of large variations in $\displaystyle \lambda_{\mbox{\tiny max}}$ (panel (e)) goes together with droplet misalignment with respect to the straining direction of the flow~\cite{Tjahjadi1991}: for such events, e.g., the one shown at around $t/\tau_{\mbox{\tiny eddy}} \approx 0.15$ in panel (f), the value of $\displaystyle\beta$ is shown to locally reach negative values, suggesting the inability of the droplet to adapt with the fast varying strain rates.
A 3D visualization of the IBLB simulation reported in panels (e-f) is provided as Supplemental Material~\cite{SupplementalMaterial}.

To delve deeper into the analysis, we performed a statistical characterization of the PDF (probability density function) of both $\displaystyle \lambda_{\mbox{\tiny max}}$ and $\displaystyle\beta$. To this aim, we considered 1000 independent turbulent trajectories~\cite{biferale2023turb} and performed IBLB numerical simulations for the values of $R/\eta=0.472$ and $\Ca=0.05, 0.15, 0.25$. Along the same trajectories, we also integrated the linear theory derived from the Stokes equations (cf. Eq.~\eqref{eq:LT}) and extracted the corresponding statistics. The two statistics are compared in  Fig.~\ref{Fig:PDF_results}. Data are displayed on both a linear scale (panels (a,c)) and a logarithmic scale (panels (b,d)). In a linear scale (panel (a)), the PDF of $\displaystyle \lambda_{\mbox{\tiny max}}/R^2$ shows a good agreement between IBLB results and the results of the linear theory for $\Ca \rightarrow 0$; at increasing $\Ca$, the IBLB PDF departs from linear theory predictions. In panel (b), the tails of the PDF of $\displaystyle \lambda_{\mbox{\tiny max}}/R^2$ are analyzed, \corr{showing that the linear theory underestimates the tail of the 
distribution and that this underestimation is more pronounced at large 
deformations, i.e., large $\Ca$.}
More statistics are probably needed for a more quantitative assessment of the distribution tails. In panel (c), we report the PDF of $\displaystyle\beta$, and we observe that the IBLB simulation results well converge to the linear theory predictions when $\Ca$ is small. The same holds for the PDF tails in panel (d). We notice that within the framework of linear theory, results for deformation are more sensitive to a variation in $\Ca$ (panels (a,b)) in comparison to the results for orientation (panels (c,d)). This is not unexpected if one thinks of the linear theory stationary results for deformation and orientation in simple flows~\cite{taylor1934formation}, where the deformation changes more than the orientation angle between the droplet and the flow direction. Overall, Fig.~\ref{Fig:PDF_results} well supports the statement that results from IBLB simulations match the linear theory predictions in the limit $\Ca \rightarrow 0$.

%%%%%%%%%%%%%%%%%%%%%%%%%%%%%%%%%%%%%%%%%%%
\section{\label{sec:conclusions}Conclusions}
%%%%%%%%%%%%%%%%%%%%%%%%%%%%%%%%%%%%%%%%%%%%%
We have investigated the application of the hybrid immersed boundary-lattice Boltzmann (IBLB) method for the simulation of the dynamics of a droplet subjected to a time-dependent velocity gradient. We have specialized the method to simulate the dynamics of a droplet in a turbulent flow, with the size of the droplet being smaller than the Kolmogorov scale of the turbulent flow. The proposed methodology hinges on two essential ingredients: first, the implementation of a generalized Zou-He scheme (cf. Sec.~\ref{sec:IBLB+BC}) to accommodate a generic strain matrix at the boundaries of the IBLB simulation domain; second, an extended Laplacian smoothing technique (cf. Sec.~\ref{subsec: regularization}) to avoid deterioration of the triangular mesh resulting in numerical instabilities. To verify the correctness of the method, we compared the results of numerical simulations with the results of the linear theory that can be \corr{\sout{analytically}} derived from the Stokes equations in the limit of small droplet deformations (i.e., small capillary numbers $\Ca$). Specifically, we have shown how the statistics of droplet deformation and orientation estimated over thousands of eddy turn-over times of the turbulent flow well match the results of the theory in the limit of small deformations. 

Various future perspectives will open up at this stage. In this paper, we mainly focused on sub-Kolmogorov droplet dynamics and relatively small values of $\Ca$. The assumption of sub-Kolmogorov droplet dynamics is crucial to have some reference theory to compare with, but in principle the LB method can be pushed towards situations where inertial contributions appear. Moreover, simulations with larger $\Ca$ are possible, but larger deformations and droplet breakup need to be suitably accounted for within the numerical simulations~\cite{cristini2003turbulence,biferale2014deformation,Ray2018} to allow a detailed characterization of breakup statistics. We also notice that we focused on the dynamics of simple droplets with surface tension at the interface. The IBLB method is flexible enough to allow the adaptation of additional interfacial complexities to model soft suspensions like vesicles or capsules, realistically employed in industrial and medical devices~\cite{art:kruger14deformability,muller2023predicting,Owen2023,de2024fluid}. This potentiality could shed light on the dynamics of such soft suspensions in time-dependent flows. Finally, we also remark that the deformation of soft suspensions in complex flows has also been investigated with phenomenological models~\cite{maffettone1998equation,minale2010models,art:arora04,Dirkes2024} based on the idea that the suspended particle retains an ellipsoidal shape at all times. Earlier studies used these models also to investigate the dynamics of droplets in turbulent flows~\cite{biferale2014deformation,Spandan2016,Ray2018}: while the predictions in the small deformation limit coincide with the linear theory that we have used in this paper, deviations are expected to emerge at large $\Ca$. It could then be of interest to use the IBLB method to design improved phenomenological models better working at large $\Ca$, in the same spirit of Ref.~\cite{taglienti2023reduced} for droplets in simple shear flow.

\subsection*{Acknowledgments}
We wish to acknowledge Fabio Bonaccorso for support. We also acknowledge useful discussion with Luca Biferale, Michele Buzzicotti and Chiara Calascibetta. This work received funding from the European Research Council (ERC) under the European Union's Horizon 2020 research and innovation programme (grant agreement No 882340). This work was supported by the
Italian Ministry of University and Research (MUR) under the FARE programme (No. R2045J8XAW), project “Smart-HEART”. 
MS acknowledges the support of the National Center for HPC, Big Data and Quantum Computing,  Project CN\_00000013 - CUP E83C22003230001,   Mission 4 Component 2 Investment 1.4, funded by the European Union - NextGenerationEU.

\onecolumn{\printbibliography}
%\printbibliography
\end{document}